 \documentclass[prb,twocolumn,showpacs,preprintnumbers,amsmath,amssymb,superscriptaddress]{revtex4}

\usepackage{graphicx}
\usepackage{dcolumn}
\usepackage{bm}

\newcommand{\kp}{{\ensuremath{\mathbf k}_\parallel}}

\begin{document}

\preprint{}

\title{Implementation of a non-equilibrium Green's function method to calculate spin-transfer torque}

\author{Christian Heiliger}%
 \email{christian.heiliger@nist.gov}
\affiliation{%
Center for Nanoscale Science and Technology, National Institute of Standards and Technology, Gaithersburg, MD 20899-6202
}
\affiliation{%
Maryland NanoCenter, University of Maryland, College Park, MD, 20742
}
\author{Michael Czerner}%
\author{Bogdan Yu. Yavorsky}%
\author{Ingrid Mertig}%
\affiliation{%
Department of Physics, Martin Luther University Halle-Wittenberg,
D-06099 Halle, Germany
}
\author{Mark D. Stiles}%
\affiliation{%
Center for Nanoscale Science and Technology, National Institute of Standards and Technology, Gaithersburg, MD 20899-6202
}


\date{\today}

\begin{abstract}
We present an implementation of the steady state Keldysh approach in
a Green's function multiple scattering scheme to calculate the
non-equilibrium spin density. This density is used to obtain the spin-transfer torque in junctions showing the magnetoresistance effect. 
We use our implementation to study the spin-transfer torque in
metallic Co/Cu/Co junctions. 
\end{abstract}

\pacs{Valid PACS appear here}
\maketitle

\section{introduction}
The discovery of the giant magnetoresistance (GMR) effect in metallic spin
valves systems\cite{baibich88,binasch89} has led to substantial research in the
field of spintronics due to the possible applications
including read heads in hard disks, storage elements in magnetic random
access memory (MRAM), and sensors. 

An effective method for writing information
into the elements is necessary for the application as storage elements in MRAM.
In particular, one has to be able to change the
magnetic orientation of the ferromagnetic leads relative to each other. One
promising approach is the current induced switching proposed by
Slonczewski \cite{slonczewski96} and by Berger.\cite{berger96} A current is driven through the
junction and becomes spin polarized in one ferromagnetic
lead. This polarization is conserved going through the spacer
layer. When the corresponding angular momentum of the polarized current is
not exactly aligned to the magnetization of the second ferromagnetic
lead, the electrons precess around the magnetic
moment of the second magnetic layer. In turn this precession leads to a
torque acting on this magnetization forcing it 
to rotate. When the current is large enough, the
magnetic orientation in the second layer can be switched. There is
also a torque acting on the first ferromagnetic layer, but this layer
is magnetically pinned.



In this paper we present an \textit{ab initio} calculation of the spin-transfer torque using a multiple scattering Green's function scheme. In
particular, the non-equilibrium spin density is calculated using the
steady state Keldysh approach (see Sec.~\ref{keldysh}). This spin
density is used to calculate the torque acting on the ferromagnetic
layer (see Sec.~\ref{torque}). We conclude by testing our approach
through an
application to a Co/Cu/Co system, which has been
studied by other authors.\cite{haney07,edwards05}

\section{method}

\subsection{Non-equilibrium spin density} \label{keldysh}
\begin{figure}
\includegraphics[width=0.95 \linewidth]{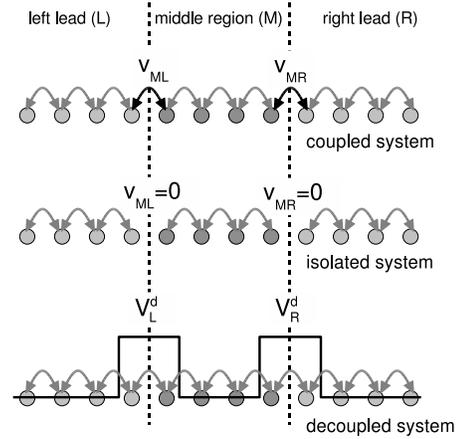}
\caption{
Division of the junction into three regions. Top: coupled (c)
system. Middle: isolated (i) system where the couplings between the middle
region and the leads are set to zero. Bottom: decoupled (d) system
where the decoupling is achieved by introducing the potentials
$V^{d}_L$ and $V^d_R$.
}
\label{fig_sketch}
\end{figure}

The non-equilibrium Green's function (NEGF) approach is based on
dividing the junction into three regions (see Fig.~\ref{fig_sketch}): two semi-infinite unperturbed
leads left (L) and
right (R) and a middle (M) region (or scattering region). 
This division allows a description of the effect on the middle region
of the semi-inifinite leads each having a different chemical
potential. The effect on the middle region can be written in terms of
a self-energy of the left lead $\Sigma_L$ given by the coupling from
the middle to the left lead and back 
\begin{equation} \label{def_sigma}
  \Sigma_L=v_{ML}\ g_L\ v^\dagger_{ML}
\end{equation}
where $g_L$ is the surface Green's function of the isolated
semi-infinite left lead and $v_{ML}$ describes the coupling of the
left lead to the middle region. In an analogous way one defines the
self-energy of the right lead $\Sigma_R$. These self-energies can
be interpreted as fluxes of incoming and outgoing electrons at
the connection between leads and middle region.\cite{datta99} Using the self-energy
of the left and right lead one can express the spin density matrix in the
middle region
\begin{equation} \label{sigma_dist}
  \rho_M= \frac{i}{2 \pi} G_{M,c} \ \left [  \left ( \Sigma_L - \Sigma^\dagger_L \right )
  f_L  + \left ( \Sigma_R - \Sigma^\dagger_R \right )
  f_R \right ] G_{M,c}^\dagger
\end{equation}
where $G_{M,c}$ is the Green's function
of region M  coupled (c) to
the semi-infinite leads and $f_{L}$ and $f_R$ are the the distribution
functions of the left and right lead.\cite{datta99} All quantities in
Eq.~\ref{sigma_dist} are energy dependent. The self-energy can be used to
relate the coupled (c) and isolated (i) Green's functions of the
middle region via a Dyson equation
\begin{equation} \label{dyson_keld}
  G_{M,c}=G_{M,i}+G_{M,i} \left( \Sigma_L + \Sigma_R  \right )
  G_{M,c} \ .
\end{equation}
This equation is based on the assumption that the self-energies can
be added, which is true if the leads are well separated and their
interaction can be neglected.

Our approach for calculating the self-energy with a \hbox{Korringa} {Kohn}
{Rostoker} (KKR) multiple scattering scheme is based on the approach
due to Henk \textit{et al.} \cite{henk06}, following the work by Pendry
\textit{et al.} \cite{pendry91} Here we sketch this idea and highlight the basic
assumptions which are necessary. 

For the calculation of the self-energy using Eq.~(\ref{def_sigma}), one
needs to know the coupling between the lead and the middle region. In
contrast to, \textit{e.g.}, tight binding approaches\cite{edwards05} this coupling is not
directly accessible within the KKR scheme because one
calculates the Green's function of a system by a Dyson equation.
Hence to obtain the coupling, one has to invert the Green's
 function.
The alternate approach we use is to introduce decoupling
potentials $V^{d}_L$ and $V^d_R$ (see Fig.~\ref{fig_sketch} bottom) which
decouple the middle region from the leads using finite barriers.
In the following we show that 
one can define a self-energy using $V^{d}$ and the
Green's function $G_{d}$ of the decoupled system.  
The Green's functions of the infinite systems (coupled
and decoupled) can be related by the Dyson equation
\begin{equation}
  G_c=G_{d} - G_{d} \left ( V^{d}_L + V^{d}_R  \right ) G_c \ .
\end{equation}
Inserting this equation once in itself and using the assumptions
(written schematically)
\begin{eqnarray} \label{assum1}
  G_{d} \left ( V^{d}_L+V^{d}_R \right ) G_{d} \ll \ \ \ \ \ \ \ \
  \ \ \ \ \ \ \ \ \ \ \ \ \ \ \ \ \ \ \ \nonumber \\
G_{d} \left (
  V_L^{d} + V^{d}_R \right) G_{d} \left (  V_L^{d}+V_R^{d}
  \right ) G_c
\end{eqnarray}
and
\begin{eqnarray} \label{assum2}
  G_{d} \left ( V^{d}_{L} G_{d} V^{d}_{L} + V^{d}_{R} G_{d}
  V^{d}_{R} \right ) G_c \gg \ \ \ \ \ \ \ \ \ \ \ \ \ \  \nonumber \\ 
  G_{d} \left (
  V^{d}_{L} G_{d} V^{d}_{R} + V^{d}_{R} G_{d} V^{d}_{L}  \right ) G_c
\end{eqnarray}
one can identify the self-energies by comparing the result to Eqs. (\ref{def_sigma}) and (\ref{dyson_keld})
\begin{equation} \label{sigma}
  \Sigma_{L}=V^{d}_{L}\ G_{d}\ \ V^{d}_{L} \ \ \text{and} \ \Sigma_{R}=V^{d}_{R}\ G_{d}\ V^{d}_{R} .
\end{equation}
Assumptions (\ref{assum1}) and (\ref{assum2}) are necessary because
$V^{d}_L$ and $V^d_R$ are local
potentials whereas $v_{LM}$ in Eq. (\ref{def_sigma}) is a coupling. Assumption (\ref{assum1}) is
fulfilled if $\left ( V^{d}_L+V^d_R \right ) G_c >> 1$ which one assures by choosing an
appropriately high $V^{d}$. Assumption (\ref{assum2}) is that the self-energies of the right and left lead can be added. This is
fulfilled if the leads are well separated because the elements
of $G_{d}$ relating the left and the right lead decay exponentially
with respect to the thickness of the decoupling potential. 

By comparing Eqs.~(\ref{def_sigma}) and (\ref{sigma}) the role of $G_{d}$ is the role of the
surface Green's function of the isolated leads. Therefore, one can also
set the whole middle region to the potential $V^{d}$ when calculating $G_{d}$.
For details of the implementation using the KKR basis set see 
Ref.~\onlinecite{henk06}. For the present work, we generalize the method
to non-collinear magnetizations based on Ref.~\onlinecite{yavorski06}.



\subsection{Spin-transfer torque} \label{torque}
From the non-equilibrium Green's function for a non-collinear
magnetization described in the previous section, it is straightforward
to compute the spin-tranfer torque. 
In linear response, the spin torque $\vec{\tau}$ per current $I$ on layer $i$ can be
expressed by (see Haney \textit{et al.} \cite{haney07})
\begin{equation}
  \frac{\vec{\tau}}{I}=2 \pi \frac{\mu_B}{e} \frac{\int d\kp \sum_l \vec{\Delta}_l
  \times \vec{m}^{tr}_l(\kp) }{\int d\kp T(\kp)}
\end{equation}
where $\vec{\Delta_l}$ are the matrix elements of the exchange field along the magnetization axis of
the layer $i$ expanded into spherical harmonics with $l$ being the
angular momentum.  $\vec{m}^{tr}_l$ is the magnetic moment of the electrons
contributing to the transport and is calculated from the
non-equilibrium spin density matrix at the Fermi level $E_F$ using
Eq.~(\ref{sigma_dist})
\begin{equation} \label{eq_den}
  \rho^{tr}=\frac{i}{2 \pi} G_c(E_F) \left (  \Sigma_L(E_F) - \Sigma_L^\dagger(E_F)
  \right ) G_c^\dagger(E_F)
\end{equation}
taking into account that the electronic states at $E_F$ are occupied
only in one lead.
The transmission probability $T$ is calculated by
\begin{equation} \label{eq_trans}
  T = \text{Tr} \left [ \left( \Sigma_L-\Sigma_L^\dagger \right) G \left(
  \Sigma_R-\Sigma_R^\dagger \right)  G^\dagger  \right] \ ,
\end{equation}
where the trace is over the spin index and the basis set expansion.
Due to the in-plane translational invariance of the junctions one can
label the states by the wave vector $\kp$ and all quantities in
Eqs.~(\ref{eq_den}) and (\ref{eq_trans}) depend on $\kp$. To get the total values one
has to integrate over the 2D Brillouin zone.


\section{Application to $\text{Co/Cu/Co}$}
\begin{figure}
\includegraphics[width=0.85 \linewidth]{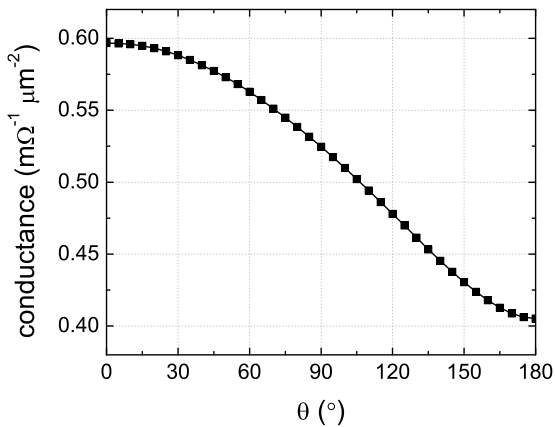}
\caption{
Conductance of the Co/Cu/Co/Cu spin valve as a function of the
relative angle between both magnetization of the ferromagnetic layers.
}
\label{fig_conduct}
\end{figure}

In this section we test our approach by applying it to a Co/Cu/Co system. In particular, we
consider the dependence of 
the torque on the angle between
the two magnetizations of the ferromagnetic leads. For this purpose,
we use the same structure used in Ref.~\onlinecite{haney07} consisting of a semi-infinite Co, 9
monolayers Cu, 15 monolayers Co, and semi-infinite Cu. The lattice
constant is 0.361~nm. Fig.~\ref{fig_conduct} shows the dependence of
the conductance $g$ on the relative
angle $\theta$ between the magnetizations of the Co layers.
 We find excellent
agreement with Ref.~\onlinecite{haney07} and a GMR ratio $\left (
g(0^\circ)-g(180^\circ) \right) / g(180^\circ)=47\ \%$.

\begin{figure}
\includegraphics[width=0.85 \linewidth]{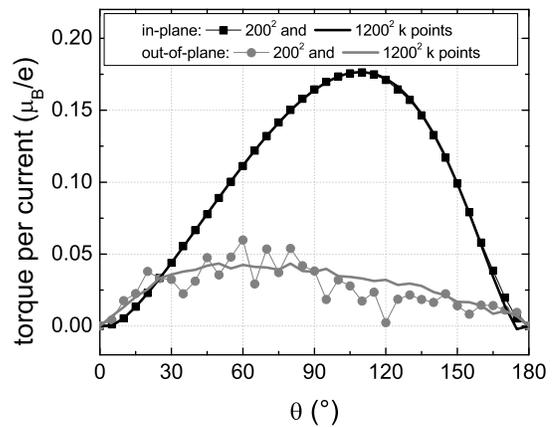}
\caption{
Torque per current as a function of the
relative angle between both magnetizations of the ferromagnetic layers
for two different numbers of $\kp$ points in the whole 2D Brillouin zone.
}
\label{fig_torque}
\end{figure}

Fig.~\ref{fig_torque} shows the in-plane torque $\tau_\parallel$ and
the out-of-plane torque $\tau_\perp$ as a
function of the angle $\theta$ for two different $\kp$
point samples.
The out-of-plane torque has two contributions: one from the right going states that are occupied and have no left going counterparts, and the other from states below both chemical potentials in which both left and right going states are occupied. The latter contribution requires integration over energy as well as parallel wave vector.  However, the usage of a complex energy contour makes it easier to converge. For the former contribution one can not use a complex energy because only right going states are occupied. Therefore, to test our method against previous calculations and to test the k-point convergence, we consider only this contribution to the out-of-plane torque in the following. 
The dependence of $\tau_\parallel$ on $\theta$ is the same for both
$\kp$ samplings and is in good agreement to Ref.~\onlinecite{haney07}. In
contrast, $\tau_\perp$ shows rapid oscillations as a function of
$\theta$ for the $\kp$ sampling using 40 000 $\kp$ points. A similar
dependence was observed in Ref.~\onlinecite{haney07}. However, a
significantly larger $\kp$ point sample leads to an almost
smooth dependence of $\tau_\perp$ on $\theta$. 

\begin{figure}
\includegraphics[width=0.85 \linewidth]{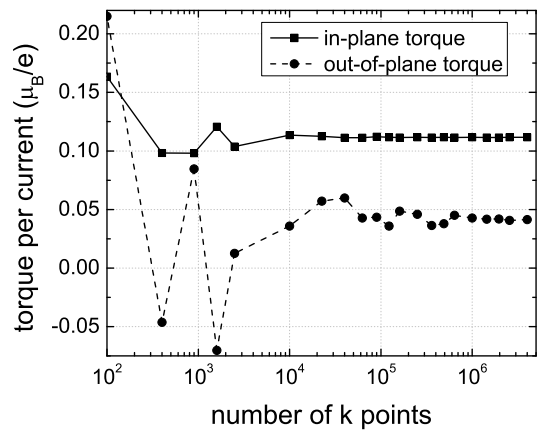}
\caption{
Convergence test for the torque with respect to the number
of $\kp$ points in the whole 2D Brillouin zone. 
}
\label{fig_conv}
\end{figure}

Due to the drastic change in $\tau_\perp$ from increasing the number of
$\kp$ points, we test the convergence of the in-plane torque $\tau_\parallel$ and
the out-of-plane torque $\tau_\perp$ for a fixed
angle $\theta=60^\circ$ as a function of the number of $\kp$ points
(see Fig.~\ref{fig_conv}). 
For $\tau_\parallel$ the convergence is fast and a relatively low number
of $\kp$ points is sufficient. On the other hand 
$\tau_\perp$ is very sensitive to the number of $\kp$  points
and a large number is necessary to get convergence. The slow
convergence results from the presence of 
short period oscillations at the corners of the 2D
Brillouin zone which require  a very fine mesh. Therefore,
the rapid oscillations in $\tau_\perp$ as a function of
angle found in Ref.~\onlinecite{haney07} disappear for fully converged
$\kp$ point samples.



\section{conclusion}
We present a method to calculate the spin-transfer torque
within a screened KKR scheme by calculating the non-equilibrium spin
density using the steady state Keldysh approach. The in-plane torque
in the Co/Cu/Co junctions is robust with respect to the $\kp$ point
sampling but the out-of-plane torque converges slowly with
respect to the number of $\kp$ points. The reason is that there are
short period oscillations at the edges of the 2D Brillouin
zone. These oscillation require a very fine $\kp$ point mesh to get the
correct value for the integral. Then both components of the torque are
a smooth
function of the angle but with maxima at different angles.

\section{Acknowledgements}
We thank J. Henk and P.M. Haney for fruitful discussions. 
This work has been supported in part by the NIST-CNST/UMD-NanoCenter
Cooperative Agreement.


\end{document}